# Determination of Intrinsic Ferroelectric Polarization in Orthorhombic Manganites with E-type Spin Order

Y. S. Chai,[1] Y. S. Oh,[1] L. J. Wang,[2] N. Manivannan,[1] S. M. Feng,[2] Y. S. Yang,[1] L. Q. Yan,[1] C. Q. Jin,[2] and Kee Hoon Kim[1],*

[1]CeNSCMR, Department of Physics and Astronomy, Seoul National University, Seoul 151-742, Republic of Korea

[2]Institute of Physics, Chinese Academy of Science, P. O. Box 603, Beijing 100080, P. R. China

By directly measuring electrical hysteresis loops using the Positive-Up Negative-Down (PUND) method, we accurately determined the remanent ferroelectric polarization $P_r$ of orthorhombic $R$MnO$_3$ ($R$ = Ho, Tm, Yb, and Lu) compounds below their E-type spin ordering temperatures. We found that LuMnO$_3$ has the largest $P_r$ of 0.17 $\mu$C/cm$^2$ at 6 K in the series, indicating that its single-crystal form can produce a $P_r$ of at least 0.6 $\mu$C/cm$^2$ at 0 K. Furthermore, at a fixed temperature, $P_r$ decreases systematically with increasing rare earth ion radius from $R$ = Lu to Ho, exhibiting a strong correlation with the variations in the in-plane Mn–O–Mn bond angle and Mn–O distances. Our experimental results suggest that the contribution of the Mn $t_{2g}$ orbitals dominates the ferroelectric polarization.

*khkim@phya.snu.ac.kr

Recent intensive researches on multiferroic materials are motivated by great interests in the fundamental physics of spin-lattice coupling as well as the potential for using these materials in the multifunctional memories and sensors.[1-4] An interesting class of multiferroic materials that have been studied intensively in recent years is the so-called magnetic ferroelectrics. In these materials, the ferroelectric polarization ($P$), induced by the primary magnetic order, can be sensitively tuned by magnetic fields through the control of magnetic states.[5] It is now well-known that both collinear and non-collinear spin orderings in these magnetic ferroelectrics can generate a nontrivial $P$ through several mechanisms such as exchange-striction[6,7] and the inverse Dzyaloshinskii–Moriya interaction.[8,9] Limited by the rather weak spin-lattice coupling strength set by these mechanisms, most of the magnetic ferroelectrics studied so far exhibit $P$ values less than 0.1 $\mu C/cm^2$, which is much smaller than that of conventional ferroelectrics.

On the other hand, several theoretical works have suggested the possibility of achieving large $P$ in orthorhombic ($o$)-$R$MnO$_3$ ($R$ = Ho, Er, Tm, Yb, and Lu) with a collinear E-type antiferromagnetic (AFM) spin order.[10-13] In this system, $Mn^{3+}$ has a $t_{2g}^3 e_g^1$ configuration and thus undergoes an $e_g$ orbital ordering with the Jahn–Teller distortion of the MnO$_6$ octahedra, leading to a distribution of long and short Mn–O bond lengths ($d_l$ and $d_s$) in the $ab$-plane [Fig. 1(a)].[14] The E-type AFM order at low temperatures is predicted to induce $P$ by two different mechanisms. First, the ionic displacement polarization $P_{ion}$ is induced by the competition between the ferromagnetic super-exchange interaction between $e_g$ orbitals and the AFM interaction between $t_{2g}$ orbitals through the inverse Goodenough–Kanamori rules,[10,11] as illustrated in Fig. 1(b). Second, the electronic polarization $P_{ele}$ results from selective electron hopping between orbitals with parallel spins and has contributions from both $t_{2g}$ and $e_g$ orbitals as well as oxygen, although the $t_{2g}$ and $e_g$ orbital contributions nearly cancel each other, leaving just mainly the O contribution in the end. As summarized in Fig. 1(b),[10] the $t_{2g}$ and $e_g$ contributions in both mechanisms have opposite signs and the total $P$ results from the sum of those contributions. Both model Hamiltonian[12] and first-principles[10] calculations predicted that a total $P$ up to approximately 6 $\mu C/cm^2$ can be obtained in $o$-HoMnO$_3$ and the value is almost same over the $o$-$R$MnO$_3$ series (from $R$ = Ho to Lu). In a subsequent study using the hybrid functional approach, the predicted value decreased to approximately 2 $\mu C/cm^2$. This reduction occurred because the hybrid functional method reduces the electronic contribution.[15]

An experimental test of these theoretical results is an important and necessary step in multiferroic research as it can not only identify the maximum allowed $P$ when the E-type spin order is present but may also help find yet another new compound generating large $P$. However, previous experimental studies on the $o$-$R$MnO$_3$ compounds have shown inconsistent results. Early studies on $o$-HoMnO$_3$ and $o$-TmMnO$_3$ revealed $P$ values inside the E-type AFM phase of approximately 0.008 $\mu C/cm^2$ at 5 K and 0.15 $\mu C/cm^2$ at 2 K.[17,18] More recently, $P$ values of 0.07–0.09 $\mu C/cm^2$ were observed at 2 K for all $o$-$R$MnO$_3$ ($R$ = Ho, Tm, Yb, Lu, and Y$_{1-y}$Lu$_y$, $y$ = 0-1), which changed little with variation in the rare earth ion radius ($r_R$).[19] It should be noted that all of these previous reports employed the pyroelectric current ($J_p$) measurement. In this method, the temperature-dependent $J_p$ is measured upon warming after applying dc electric poling from a high temperature above the ferroelectric Curie temperature to the low temperature at which the measurement starts. The temperature dependent polarization $P_{dc}(T)$ can be attained by integrating $J_p$ as a function of time. However, the common procedures followed in this method turned out to provide inaccurate $P$ data due to several experimental challenges. First, most of the $o$-$R$MnO$_3$ ($R$ =

Ho to Lu) compounds have a polycrystalline pellet form synthesized under a high pressure. Because of this, the $J_p$ measurement is subject to incomplete electric poling and thus the ferroelectric domains can be randomly oriented. Second, the electric poling process usually produces space charges that can be trapped at the polycrystalline grain boundaries, providing a spurious $P$ contribution. These fundamental challenges in determining the $P$ value in polycrystalline ferroelectrics can be greatly reduced by employing the so-called Positive-Up Negative-Down (PUND) method,[20] which has been extensively employed in ferroelectric thin film researches and recently has been applied to multiferroic single crystals.[21] In particular, we have recently used this technique to prove that a polycrystalline $o$-HoMnO$_3$ sample has an intrinsic $P$ of approximately 0.24 $\mu$C/cm$^2$,[22] which is much smaller than that predicted by band calculations.[10,12,15]

In this communication, we report our systematic efforts to determine the intrinsic ferroelectric polarization of $o$-$R$MnO$_3$ for $R$ = Ho, Tm, Yb, and Lu with E-type spin order. By applying the PUND method to high quality specimens that have been well characterized by structural studies, we succeeded in measuring electrical hysteresis loops for all the samples as a function of temperature. We found that the remanent ferroelectric polarization $P_r$ systematically increases with decreasing $r_R$ and thus $o$-LuMnO$_3$ (HoMnO$_3$) produces the largest (smallest) $P_r$ value of 0.17 (0.068) $\mu$C/cm$^2$ at 6 K. Based on the local structure analysis, we suggest that the orthorhombic manganites with E-type spin order have higher electronic polarization contributions from the $t_{2g}$ than the $e_g$ orbitals.

$o$-$R$MnO$_3$ ($R$ = Ho, Er, Tm, Yb, and Lu) specimens were synthesized under a high pressure of 5 GPa at 1423 K[23] and an X-ray diffraction study confirmed the orthorhombic (*Pbnm*) structure at 300 K was free of impurities. Crystal structures were extracted by Rietveld refinement using the GSAS program. All the specimens investigated for the hysteresis loop had a density at least 95% of their theoretical value. We made thin plate-like samples with thickness of approximately 0.3 mm, used silver epoxy (EPTEK H20E) to make electrodes. We employed a PPMS$^{TM}$ (Quantum Design) to control temperature environment for the hysteresis loop or $J_p$ measurements. For the PUND method, we applied a series of positive (P$i$, $i = 0$–2) and negative (N$i$, $i = 0$–2) electric pulses as shown in Fig. 2(a). The first two pulses, P0 and N0, are used to fully align the ferroelectric domains. During the next two pulses, P1 and P2 (N1 and N2), two curves representing effective polarization changes are recorded in the Sawyer–Tower circuit, and they are subtracted to form the half loop for electric field $E > 0$ ($E < 0$) [Fig. 2(b)]. As a result, the pure hysteretic parts of the hysteresis loop can be obtained without being obscured by resistive or capacitive components. Moreover, by employing a short pulse, the maximum peak field for electric poling can be increased to better align the ferroelectric domains without inducing electrical break-down effects. In particular, the space charge effect is minimized as the sample is poled in isothermal conditions. These features enable us to mitigate the experimental problems encountered in conventional $J_p$ measurements with dc field poling.

Figure 2(c) displays a typical electrical hysteresis loop obtained by this procedure, for the case of $o$-LuMnO$_3$ at 6 K. The y-axis offset directly represents $P_r \approx 0.17$ $\mu$C/cm$^2$. At each temperature, we have measured the loop by increasing amplitude of the pulse and then determined the polarization until the electrical break-down happens. The $P_r$ vs. maximum $E$ curves thus obtained for $o$-LuMnO$_3$ at 15 K and 25 K show almost saturation at the high field, suggesting that $P_r$ at the maximum $E$ of 11.8 MV/m is close to the intrinsic polarization due to fully aligned ferroelectric domains. To verify thus obtained $P_r$ value by the PUND method, the pyroelectric current $J_p$ has been also measured after the short N2 pulse [Fig. 2(a)], and the temperature dependence of $P$, termed as $P_{pls}(T)$ in Fig. 2(c), has been estimated. The difference between $P_r$ and $P_{pls}(T)$ turns out to be less than 5% at 6 K, supporting the conclusion that both $P_r$ and $P_{pls}(T)$ are close to the intrinsic ferroelectric polarization. However, the $P_{dc}(T)$ curves determined through the $J_p$ measurement after dc electric field poling did not show saturation before an electrical break-down. Moreover, the maximum polarization values obtained at both 15 K and 25 K were clearly larger than those obtained using the PUND method. It is most likely that the conventional $P_{dc}(T)$ measurements include significant contributions from trapped space charges that have accumulated during the poling process, while the $P_{pls}(T)$ measurements do not.

By applying the same experimental method, we determined the electrical hysteresis loops at various temperatures for all the $o$-$R$MnO$_3$ ($R$ = Ho, Tm, Yb, and Lu) after cooling the samples without an electric field bias [Figs. 3(a)–3(d)], and summarized the resultant temperature-dependence of $P_r$ and the $P_{pls}(T)$ curves [Figs. 3(e)–3(h)]. The onset temperatures of ferroelectric polarization were found to be 26, 35, 37, and 38 K for $R$ = Ho, Tm, Yb, and Lu, respectively and they are consistent with the reported lock-in transition temperature, $T_L$, in each compound.[14,18,22,23] Surprisingly, $o$-LuMnO$_3$ clearly showed the largest $P_r$ value of 0.17 $\mu$C/cm$^2$ at 6 K and upon extrapolation, $P_r$ would reach approximately 0.2 $\mu$C/cm$^2$ at 0 K [Fig. 3(d)]. Because the E-type spin order is supposed to generate uniaxial electric polarization along the $a$-axis, we can expect that a polycrystalline specimen would have roughly one third of the single crystal polarization value because of the random orientation of grains. Therefore, $P_r$ = 0.2 $\mu$C/cm$^2$ in our polycrystalline sample predicts at least $P_a$ = 0.6 $\mu$C/cm$^2$ in the $o$-LuMnO$_3$ single crystal. We note that this $P_a$ value is significantly larger than those observed in typical magnetic ferroelectrics although it is still much lower than the theoretically predicted values of approximately 2–6 $\mu$C/cm$^2$.[10,15]

It is uniquely revealed in our investigations that the ferroelectric polarization in the studied orthorhombic manganites systematically changes with the variation in $r_R$. Figure 3 shows that with the decrease in $r_R$ from $o$-HoMnO$_3$ to $o$-LuMnO$_3$, $P_r$ values increase systematically over all temperatures. Figure 4(a) summarizes the 3$P_r$ vs. $r_R$ relationship at temperatures of 6 K and $T_L$/2 (= 13, 17.5, 18.5, and 19 K for $R$ = Ho, Tm, Yb, and Lu, respectively), proving that the 3$P_r$ values increase when the $r_R$ is reduced. Upon comparing those 3$P_r$ values of $o$-HoMnO$_3$ and $o$-LuMnO$_3$, we find that the 3$P_r$ values increase by 1.2 and 0.67 times at 6 K and $T_L$/2, respectively. Because the hysteresis loops at $T_L$/2 are clearly saturated at high $E$ while the loops at 6 K are less saturated, the smoothly increasing tendency observed at $T_L$/2 should reflect the intrinsic $r_R$ dependence. We note that our observations are in contrast with the theoretical results in Ref. 10, which predicts an almost constant $P$ behavior regardless of $r_R$ changes. Furthermore, our results are also inconsistent with the experimental data in Ref. 19, in which the $J_p$ measurements subject to the trapped space charge problems resulted in $P$ values that were almost constant over different $r_R$. We also note that different annealing treatments under O$_2$, N$_2$, and air atmospheres in our $o$-HoMnO$_3$ specimen did not show any significant changes in the $P_r$ vs. temperature curves,[22] suggesting that the observed $r_R$ dependence is rather insensitive to oxygen stoichiometry in the $o$-$R$MnO$_3$ specimens. Therefore, it is likely that the current 3$P_r$ vs. $r_R$ data reflect the intrinsic polarization behavior in $o$-$R$MnO$_3$ ($R$ = Ho, Tm, Yb, and Lu).

In order to understand the origin of this rather clear change of $P_r$ with $r_R$, we determined the structural parameters of the samples at 300 K and the results are summarized in Figs. 4(b) and 4(c). The resultant lattice constants and in-plane Mn–O–Mn bond angle $\phi$ are quite similar to the reported experimental values.[23] Indeed, the $\phi$ value systematically decreases with $r_R$, showing that the compounds with a smaller $r_R$ result in a more distorted local structure. On the other hand, the Mn–O bond lengths, $d_l$ and $d_s$, clearly show opposite tendency with the variation in $r_R$; $d_l$ ($d_s$) increases (decreases) with the decrease in $r_R$ [Fig. 4(c)]. This experimental finding is consistent with other experiment results[23] but is in contrast with the behavior of the input parameters used in the first principles calculation,[10] in which the optimized crystal structure in the E-type spin order results in the decreasing behaviors for both $d_l$ and $d_s$ with decreasing $r_R$. Figure 4(c) compares these contrasting experimental and theoretical behaviors of $d_l$ and $d_s$ with $r_R$. Moreover, we note that the experimental changes of $d_l$ and $d_s$ over $r_R$, which turn out to be approximately 3%, are much bigger than the theoretical predictions, which are less than 0.5%. These lattice parameter variations over $r_R$ are thus expected to hold even at low temperatures because thermal shrinking in these $o$-$R$MnO$_3$ compounds is estimated to be less than 0.3% between 300 and 10 K.[18]

The first principles calculation in Ref. 10 discussed how the input structural parameters can crucially affect the electronic polarization $P_{ele}$, which is dominant over the ionic polarization, $P_{ion}$. Firstly, with decreasing $d_l$, the hopping integral between the $e_g$ orbitals increases and thus the $e_g$ contribution to $P_{ele}$ increases. Secondly, the decrease of $d_s$ can enhance the hopping between $t_{2g}$ orbitals so that the $t_{2g}$ contribution to $P_{ele}$ will increase too. Because the enhanced contributions of $e_g$ and $t_{2g}$ orbitals are opposite in sign, the total $P$ should eventually become almost independent of $r_R$ and then become close to 6 $\mu$C/cm$^2$ in all the $o$-$R$MnO$_3$ ($R$ = Ho to Lu).[10] The dashed lines in Fig. 4(d) schematically describe these theoretical predictions for the hopping integrals and related contributions to $P_{ele}$.

On the other hand, our new experimental results for $d_l$ and $d_s$ in Fig. 4(c) suggest a new scenario that with decreasing $r_R$, the hopping integral between $e_g$ orbitals should be suppressed significantly while that between $t_{2g}$ orbitals should be increased [solid lines in Fig. 4(d)]. Accordingly, we can expect that the $t_{2g}$ orbitals contribute to $P_{ele}$ more significantly than $e_g$ orbitals overall, and this tendency would increase more as $R$ changes from Ho to Lu [Fig. 4(d)]. As the ionic and oxygen contributions were relatively small and nearly independent of $r_R$, the total $P$ would be then enhanced in proportion to $P_{ele}$. The observed increase of $3P_r$ from $o$-HoMnO$_3$ to $o$-LuMnO$_3$ seems consistent with this qualitative explanation based on the existing theoretical prediction. It will be worth further theoretical investigation based on the structural and electrical informations provided here to see whether the existing theoretical framework is still valid or requires other explanations to understand the intrinsic polarization value and its rare earth dependence in $o$-$R$MnO$_3$.

To conclude, we determined the intrinsic ferroelectric polarization in $o$-$R$MnO$_3$ ($R$ = Ho, Tm, Yb, and Lu) with E-type spin order by using the PUND method. The obtained polarization values increase systematically upon reducing the rare earth ionic radius from $R$ = Ho to Lu, and the maximum ferroelectric polarization value at 0 K is estimated to be approximately 0.6 $\mu$C/cm$^2$ in $o$-LuMnO$_3$. Our structural analyses imply that $t_{2g}$ rather than $e_g$ orbitals play a more crucial role in determining ferroelectric polarization.

We thank Y. Liu for helpful discussion. This work was financially supported by the National Creative Research Initiative (2010-0018300) and the Fundamental R&D Program for Core Technology of Materials of MOKE.

**References**

[1] T. Kimura, N. Goto, H. Shintani, T. Arima and Y. Tokura, Nature (London) **426**, 55 (2003).

[2] N. Hur, S. Park, P. A. Sharma, J. S. Ahn, S. Guha and S.-W. Cheong, Nature (London) **429**, 392 (2004).

[3] M. Fiebig, J. Phys. D **38**, R123 (2005).

[4] D. I. Khomskii, J. Magn. Magn. Mater. **306**, 1 (2006).

[5] S.-W. Cheong and M. Mostovoy, Nature Mater. **6**, 13 (2007).

[6] L. C. Chapon, G. R. Blake, M. J. Gutmann, S. Park, N. Hur, P. G. Radaelli and S.-W. Cheong, Phys. Rev. Lett. **93**, 177402 (2004).

[7] Y. J. Choi, H. T. Yi, S. Lee, Q. Huang, V. Kiryukhin and S.–W. Cheong, Phys. Rev. Lett. **100** 047601 (2006).

[8] M. Kenzelmann, A. B. Harris, S. Jonas, C. Broholm, J. Schefer, S. B. Kim, C. L. Zhang, S.-W. Cheong, O. P. Vajk and J. W. Lynn, Phys. Rev. Lett. **95**, 087206 (2005).

[9] T. Arima, A. Tokunaga, T. Goto, H. Kimura, Y. Noda and Y. Tokura, Phys. Rev. Lett. **96**, 097202 (2006).

[10] K. Yamauchi, F. Freimuth, S. Blugel and S. Picozzi, Phys. Rev. B **78,** 014403 (2008).

[11] K. Yamauchi and S. Picozzi, J. Phys. Condens. Mat. **21**, 064203 (2009).

[12] I. A. Sergienko, C. Sen and E. Dagotto, Phys. Rev. Lett. **97**, 227204 (2006).

[13] S. Picozzi, K. Yamauchi, B. Sanyal, I. A. Sergienko and E. Dagotto, Phys. Rev. Lett. **99**, 227201 (2007).

[14] Y. Murakami, J. P. Hill, D. Gibbs, M. Blume, I. Koyama, M. Tanaka, H. Kawata, T. Arima, Y. Tokura, K. Hirota and Y. Endoh, Phys. Rev. Lett. **81**, 582 (1998).

[15] A. Stroppa and S. Picozzi, Phys. Chem. Chem. Phys. **12**, 5405 (2010).

[16] S. Picozzi, K. Yamauchi, G. Bihlmayer and S. Blügel, Phys. Rev. B **74**, 094402 (2006).

[17] B. Lorenz, Y. Q. Wang and C. W. Chu, Phys. Rev. B **76**, 104405 (2007).

[18] V. Yu. Pomjakushin, M. Kenzelmann, A. Dönni, A. B. Harris, T. Nakajima, S. Mitsuda, M. Tachibana, L. Keller, J. Mesot, H. Kitazawa and E. T.-Muromachi, New J. Phys. **11**, 043019 (2009).

[19] S. Ishiwata, Y. Kaneko, Y. Tokunaga, Y. Taguchi, T. Arima and Y. Tokura, Phys. Rev. B **81**, 100411(R) (2010).

[20] J. F. Scott, L. Kammerdiner, M. Parris, S. Traynor, V. Ottenbacher, A. Shawabkeh and W. F. Oliver, J. Appl. Phys. **64**, 787 (1988).

[21] M. Fukunaga and Y. Noda, J. Phys. Soc. Jpn. **77**, 064706 (2008).

[22] S. M. Feng, Y. S. Chai, J. L. Zhu, N. Manivannan, Y. S. Oh, L. J. Wang, Y. S. Yang, C. Q. Jin and Kee Hoon Kim, New J. Phys. **12**, 073006 (2010).

[23] M. Tachibana, T. Shimoyama, H. Kawaji, T. Atake, and E. T.-Muromachi, Phys. Rev. B **75**, 144425 (2007).

**Figure Captions:**

FIG. 1. (a) $ab$-plane arrangement of Mn and O atoms and spins in the E-type spin order realized in $o$-$R$MnO$_3$. (b) The arrows represent the O atom displacement due to $e_g$ (solid) and $t_{2g}$ (dashed) orbitals. (c) The ellipses indicate the Mn charge deviations due to the $e_g$ (along $d_l$) and $t_{2g}$ (along $d_s$) hoppings, as described in Ref. 10. Arrows at the bottom depict the resultant polarizations from each contribution categorized into $P_{ion}$ and $P_{ele}$.

FIG. 2. (a) Electrical pulse patterns used in the PUND method. (b) The schematic view to extract the hysteresis loop out of the pulse sequences in (a). (c) Typical hysteresis loop at 6 K and (d) $P_{pls}(T)$ after the N2 pulse. (e) Remanent polarization $P_r$ vs. peak electric field $E$ applied at 15 and 25 K for $o$-LuMnO$_3$. Polarization values (at 15 and 25 K) obtained by the conventional pyroelectric current measurements after dc electric field poling, $P_{dc}(T)$, are also plotted as a function of applied dc electric field.

FIG. 3. Electrical hysteresis loops of (a) $o$-HoMnO$_3$, (b) $o$-TmMnO$_3$, (c) $o$-YbMnO$_3$, and (d) $o$-LuMnO$_3$. Remanent polarization $P_r$ values and related $P_{pls}(T)$ curve are summarized in (e)–(h). The data in (e) are reproduced from Ref. 22.

FIG. 4. The rare earth ionic radius $r_R$ dependence for (a) $3P_r$ values at 6 K and $T_L/2$, (b) lattice constants and bond angle ($\phi$), and (c) experimental $d_s$ and $d_l$ (solid symbols) at room temperature and calculated average $d_s$ and $d_l$ (open symbols) (from Ref. 10). (d) Schematic diagram for expected hopping integrals for $t_{2g}$ (top panel) and $e_g$ (middle panel) orbitals, and related electronic polarization contributions (bottom panel). Solid and dash lines represent the expected contributions from experimental and calculated data.

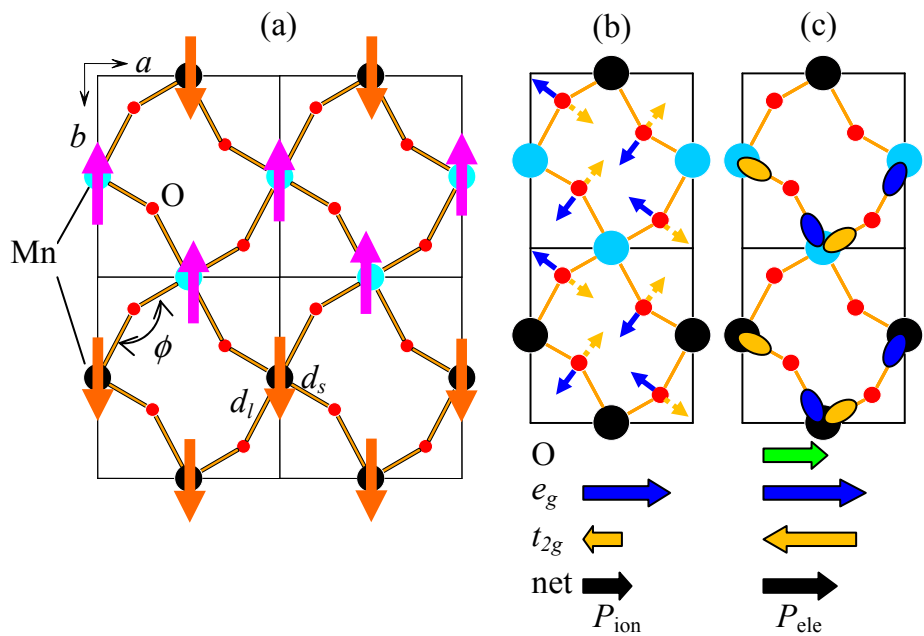

Figure 1

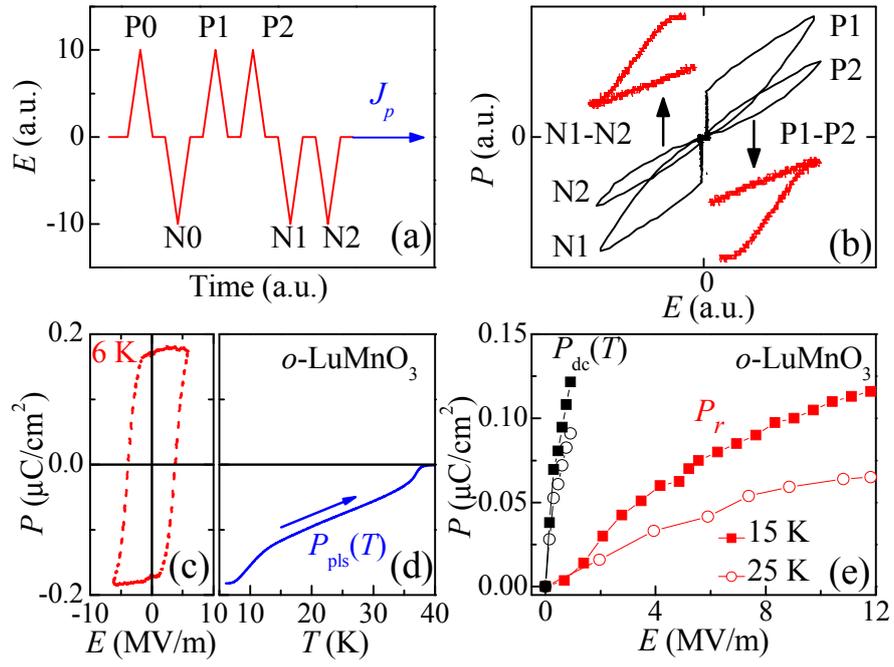

Figure 2

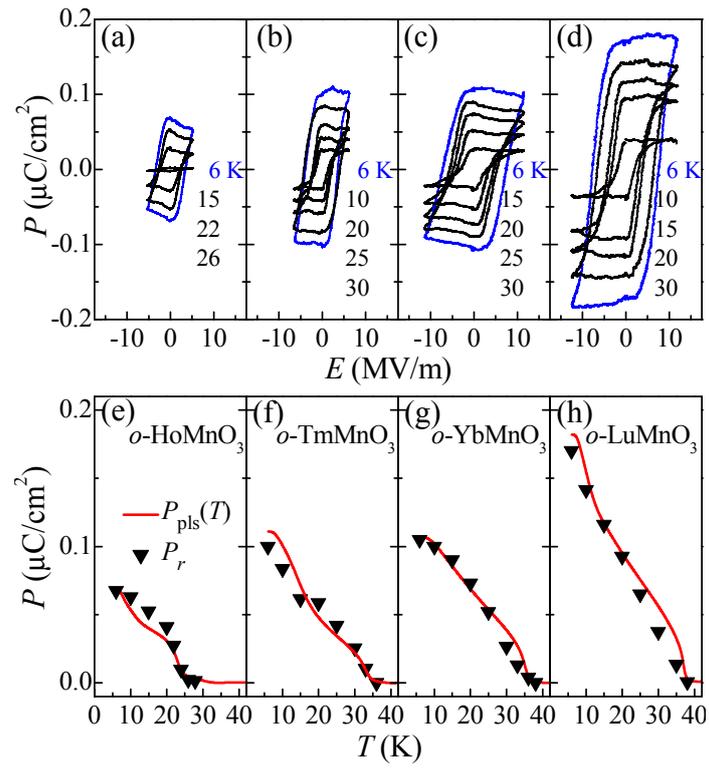

Figure 3

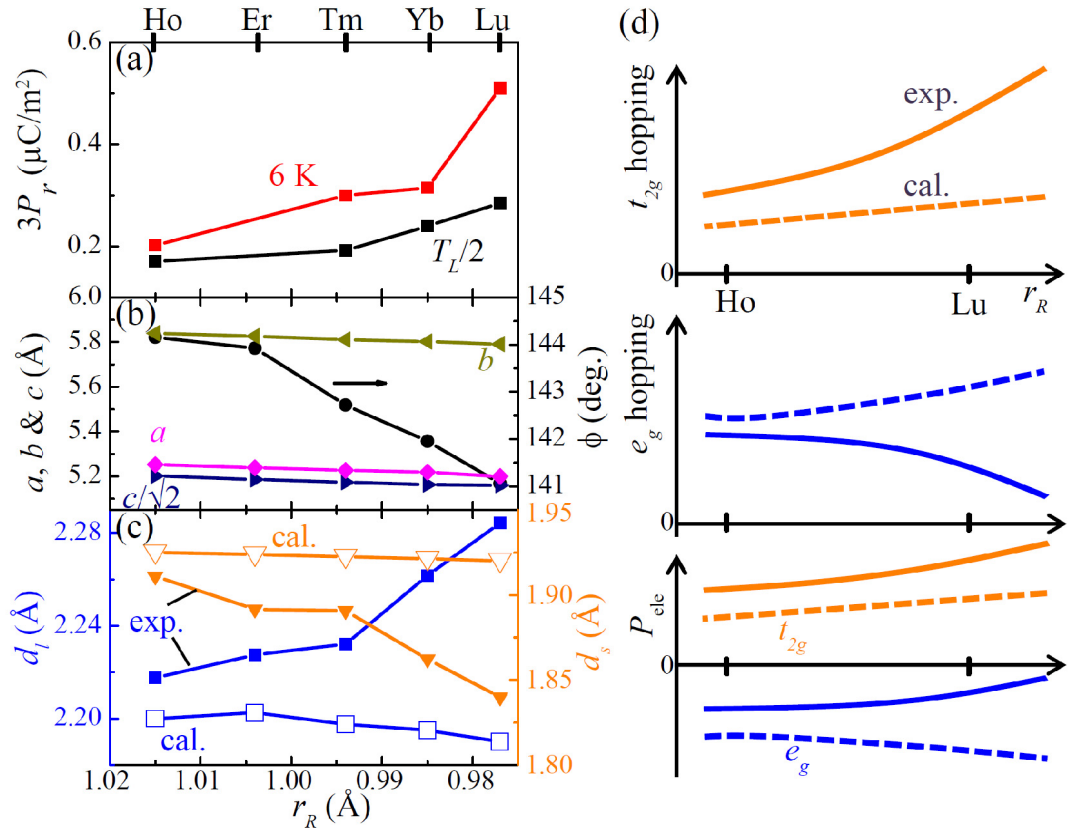

Figure 4